\documentclass[aps,prd,floatfix,nofootinbib,superscriptaddress,twocolumn,10pt]{revtex4}

\usepackage{graphicx}
\usepackage{tabularx}

\begin{document}

\title{Possibilities for methanogenic and acetogenic life in molecular cloud}
\author{Lei~Feng}
\email{fenglei@pmo.ac.cn}
\affiliation{Key Laboratory of Dark Matter and Space Astronomy, Purple Mountain Observatory, Chinese Academy of Sciences, Nanjing 210023}
\affiliation{School of Astronomy and Space Science, University of Science and Technology of China, Hefei, Anhui 230026, China}
\affiliation{Joint Center for Particle, Nuclear Physics and Cosmology,  Nanjing University -- Purple Mountain Observatory,  Nanjing  210093, China}

\begin{abstract}
According to panspermia, life on Earth may have originated from life forms transported through space from elsewhere. These life forms could have passed through molecular clouds, where the process of methanogenesis could have provided enough energy to sustain living organisms. In this study, we have calculated the Gibbs free energy released from synthesizing hydrocarbons for methanogenic (acetogenic) life in a molecular cloud, with methane (acetic acid) as the final metabolic product. Our calculations demonstrate that the chemical reactions during methanogenesis can release enough free energy to support living organisms. The methanogenic life may have served as the predecessor of life on Earth, and there is some preliminary evidence from various molecular biology studies to support this idea. Furthermore, we propose a potential distinguishing signal to test our model.
\end{abstract}
\maketitle
\section{Introduction}

In a recent paper, the authors found that the last universal common ancestor (LUCA) appeared 4.2 billion years ago, just 400 million years after the formation of Earth \cite{LUCA}. It may be a hint of panspermia. According to panspermia~\cite{Panspermia} and local panspermia \cite{nebula-relay}, life on Earth could have descended from forms of life transported through space. Such life forms could have passed through and survived within molecular clouds. Hydrogen molecules maintain a liquid state within the temperature range of 13.99~K to 20.27~K, which coincidentally coincides with the typical temperature of molecular clouds. Although it is possible to assume the liquid hydrogen environment of molecular cloud life, more severe problems arise correspondingly. So we won't discuss such a possibility here. Such an extremely low-temperature environment may have unexpected benefits. In Ref ~ \cite{Chirality}, the author argues that the ultra-low temperature environment of molecular clouds may be the reason for the chiral polymer chain of biological molecules. For more information about the effects of low temperatures on the origin of life, see Ref. \cite{lowt1,lowt2}. Additionally, the interstellar medium might harbor pre-biological compounds that foster life-sustaining processes. Although organic molecules like amino acids remain undiscovered in molecular clouds, the presence of numerous amino acids in meteorites \cite{Murchison,Engel,Cronin,Engel2}, which form in molecular clouds,  suggests the plausible existence of amino acids within molecular clouds.

A crucial question arises concerning the energy acquisition strategies employed by the cloud lifeforms. The author has previously suggested a bioenergetic mechanism driven by cosmic rays \cite{cosmic-ray-bioenergetics}, which relies on the ionization of hydrogen molecules as its primary energy source. However, alternative energy acquisition pathways that may support life in these extreme environments are worth exploring.

Metabolic activity and reproduction of biological systems require energy transformations through respiration processes, such as aerobic and anaerobic respiration. Aerobic and anaerobic respiration are all redox reactions of biological fuels in the presence of an inorganic electron acceptor. The difference between them is whether the electron acceptor is oxygen. Such redox reactions release Gibb Free energy and produce large amounts of energy stored in ATP.

On the Earth, methanogenic bacteria reduce $\rm CO_2$ to $\rm CH_4$ and release Gibbs free energy needed for life through methanogenesis, which is anaerobic respiration with methane as the final product of metabolism and only found in the domain Archaea. C.P. McKay and H.D. Smith explored the biochemical reactions of hypothetical methane-based life on Titan, with reactants including $\rm C_2H_2$, $\rm C_2H_6$, and organic haze~\cite{McKay2005}. Given the presence of similar compounds in molecular clouds, it is plausible to consider methanogens thriving in these cosmic regions as well. Here, we will discuss such probability and calculate the release of free energy for methanogenic life in the environment of molecular clouds. Since acetogenic life has the same electron donor and acceptor through the Wood-Ljungdahl pathway on Earth, the energy release of such biochemical processes in molecular clouds is also discussed in this draft. Then, we discuss the relationship between molecular cloud life and LUCA.


This paper is organized as follows: In Sec.2, we explore the possible biochemical reactions of methanogenic and acetogenic life in molecular clouds and calculate the released Gibbs free energy. A possible distinguishing signal is also presented in this section. The relation between such methanogenic and acetogenic life in molecular clouds and LUCA is discussed in Sec 3. The conclusions are summarized in the final section.

\section{Possibilities for methanogenic life in molecular cloud}





Methanogens on Earth maintain their activity through methanogenesis, a form of anaerobic respiration. Carbon dioxide is the terminal electron acceptor in methanogenesis through the following chemical reaction
\begin{equation}
\rm{CO_2+4H_2\rightarrow CH_4+2H_2O}.
\end{equation}
With the same electron acceptor and donor, acetogenic processes are also possible through the following reaction
\begin{equation}
\rm{2CO_2+4H_2\rightarrow CH_3COOH+2H_2O}.
\end{equation}
As the $\rm CO_2$ in molecular clouds is more abundant in the solid phase than in the gas phase~\cite{Minh1988,Gerakines,Whittet}, solid $\rm CO_2$ provides abundant carbon sources and energy for metabolic activities. Then, life forms may congregate around solid carbon dioxide. The triple point of carbon dioxide occurs at a temperature of 216.58 K and a pressure of 5.11 atm. At temperatures below this threshold, solid carbon dioxide can sublimate directly into a gaseous state as the temperature rises. This transition temperature notably decreases with a reduction in pressure. The temperature is sufficiently low in the place where solid carbon dioxide has been detected. Conversely, in regions where stars are forming, the temperatures are considerably higher, making it challenging to sustain solid carbon dioxide.

The most abundant molecule in molecular clouds is H2, accounting for approximately 70\% of the composition. It is also the main collision target of other molecules therein. The second most abundant molecule in the molecular cloud is $\rm CO$, and its hydrogenation reaction, as shown in Eq. \ref{CO}, cloud be the energy source of methanogen-like creatures.
\begin{equation}
\rm{CO+3H_2\rightarrow CH_4+H_2O}.
\label{CO}
\end{equation}
In addition, we also considered the chemical reaction given in Ref. \cite{McKay2005}, which is
\begin{equation}
\rm{C_2H_2+3H_2\rightarrow 2CH_4}.
\label{C2H2}
\end{equation}

\subsection{The calculation of Gibbs free energy}

The Gibbs free energy released from chemical reactions at temperature $T$ and standard 1 atm can be calculated using the free energies of formation \cite{Stull,Miller}, which is
\begin{equation}
\Delta G^\circ=\Delta H-T\Delta S,
\end{equation}
where $\Delta H$ ($\Delta S$) is the difference in heats of formation (entropy) of product and reactant under standard conditions, which is shown in Tab. \ref{molecular}.

The free energy between any pressures and standard 1 atm is given by
\begin{equation}
\Delta G=\Delta G^\circ+RT{\rm ln}(Q),
\end{equation}
where $ R$ is the universal gas constant, and $Q$ is the ratio of the activities between products and reactants raised to the power of its multiplying constant in the chemical reaction equation. Following Ref. \cite{Kral1998,McKay2005}, the chemical activities of the gas molecules we used in this draft are approximately equal to their partial pressure which is proportional to the gas mole fraction in the mixture. The activity of solid carbon dioxide is unity. For example for the $\rm C2H2$ reaction,
\begin{equation}
Q = [pCH_4]^2/[pC_2H_2][pH_2]^3.
\end{equation}

\begin{table*}[htbp]
\centering
\caption {Heats of formation and entropy at standard conditions ($\rm 25^{\circ}$C, 1 bar)}
\begin{tabularx}{\textwidth}{ccccccc}
\hline \hline
~~~~~~~~~~~~~~~~~~~~~~~~~Molecule&~~&$H({\rm kJ~mol^{-1}})$&~~&$S({\rm J/mol~K})$&~~&Abundance \\

\hline
~~~~~~~~~~~~~~~~~~~~~~~~~$\rm H_2(g)$ &~~~~~&0&~~~~~& 130.68\cite{Hydrogen}&~~~~~&$1.1\times10^4~n_{\rm CO}$\cite{Frerking_1982}\\
~~~~~~~~~~~~~~~~~~~~~~~~~$\rm CH_4(g)$ &~~~~~&-74.6\cite{Methane}&~~~~~& 186.3\cite{Methane}&~~~~~&${\rm 10^{-3}~}n_{\rm CO}$\cite{Lacy_1991}\\
~~~~~~~~~~~~~~~~~~~~~~~~~$\rm C_2H_2(g)$ &~~~~~&227.400\cite{Methane}&~~~~~& 200.927\cite{Methane}&~~~~~&${\rm 3 \times 10^{-4} \sim 10^{-3}} ~n_{\rm CO}$\cite{Lacy_1989} \\
~~~~~~~~~~~~~~~~~~~~~~~~~$\rm H_2O(g, gas)$&~~~~~&-241.83~\cite{NIST}&~~~~~& 188.84~\cite{NIST}&~~~~~&${\rm 10^{-9} \sim 10^{-7}}~n_{\rm H_2}$(for ortho-$\rm H_2O$)\cite{Snell_1999} \\
~~~~~~~~~~~~~~~~~~~~~~~~~$\rm CO(g)$ &~~~~~&-110.53~\cite{NIST}&~~~~~& 197.66~\cite{NIST}&~~~~~&  $n_{\rm CO}$\\
~~~~~~~~~~~~~~~~~~~~~~~~~$\rm CO_2(g,solid)$ &~~~~~&-427.4&~~~~~& 51.07\cite{Giauque}&~~~~~& - \\
~~~~~~~~~~~~~~~~~~~~~~~~~$\rm CH_3COOH$ &~~~~~&-433&~~~~~& 282.84\cite{Weltner}&~~~~~& ${\rm 3.4\times 10^{-10}}~n_{\rm H_2}$\cite{Mehringer} \\
  \hline
  \hline
\end{tabularx}
\label{molecular}
\end{table*}

The typical temperature of a molecular cloud is about $\rm 10\sim20~K$, but temperatures of "warm clouds" are about 20-60 K and molecular clouds with H II regions can reach 100 K~\cite{temperature}. So, we set $\rm 10~K <T< 100~K$ in this draft. The partial pressure of $\rm H_2$, $\rm CH_4$, $\rm C_2H_2$, $\rm H_2O$, $\rm CO$ and  $\rm CH_3COOH$ we used here are 70\%, $6\times10^{-8}$, $3\times10^{-8}$, $7\times10^{-9}$, $6.36\times10^{-5}$ and $2.38\times10^{-10}$, respectively. Utilizing these equations and parameters, we can compute the Gibbs free energy released for the methane and acetic acid production processes.

\subsection{The results}

\begin{figure}[htbp]
\centering
\includegraphics[width=0.5\textwidth]{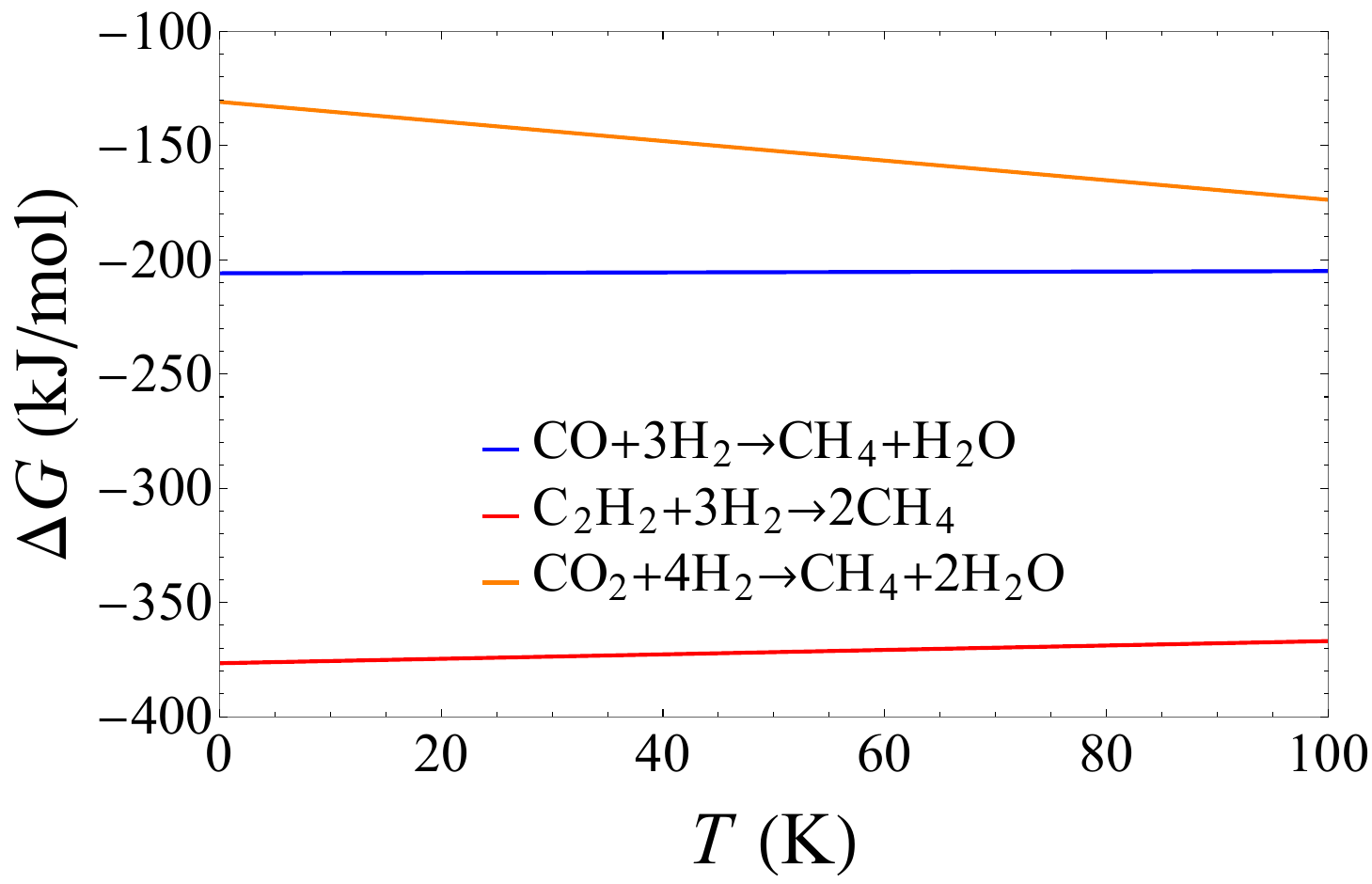}
\caption{The Gibbs free energy released from the synthesis of methane.}
\label{fig:deltaG1}
\end{figure}

\begin{figure}[htbp]
\centering
\includegraphics[width=0.5\textwidth]{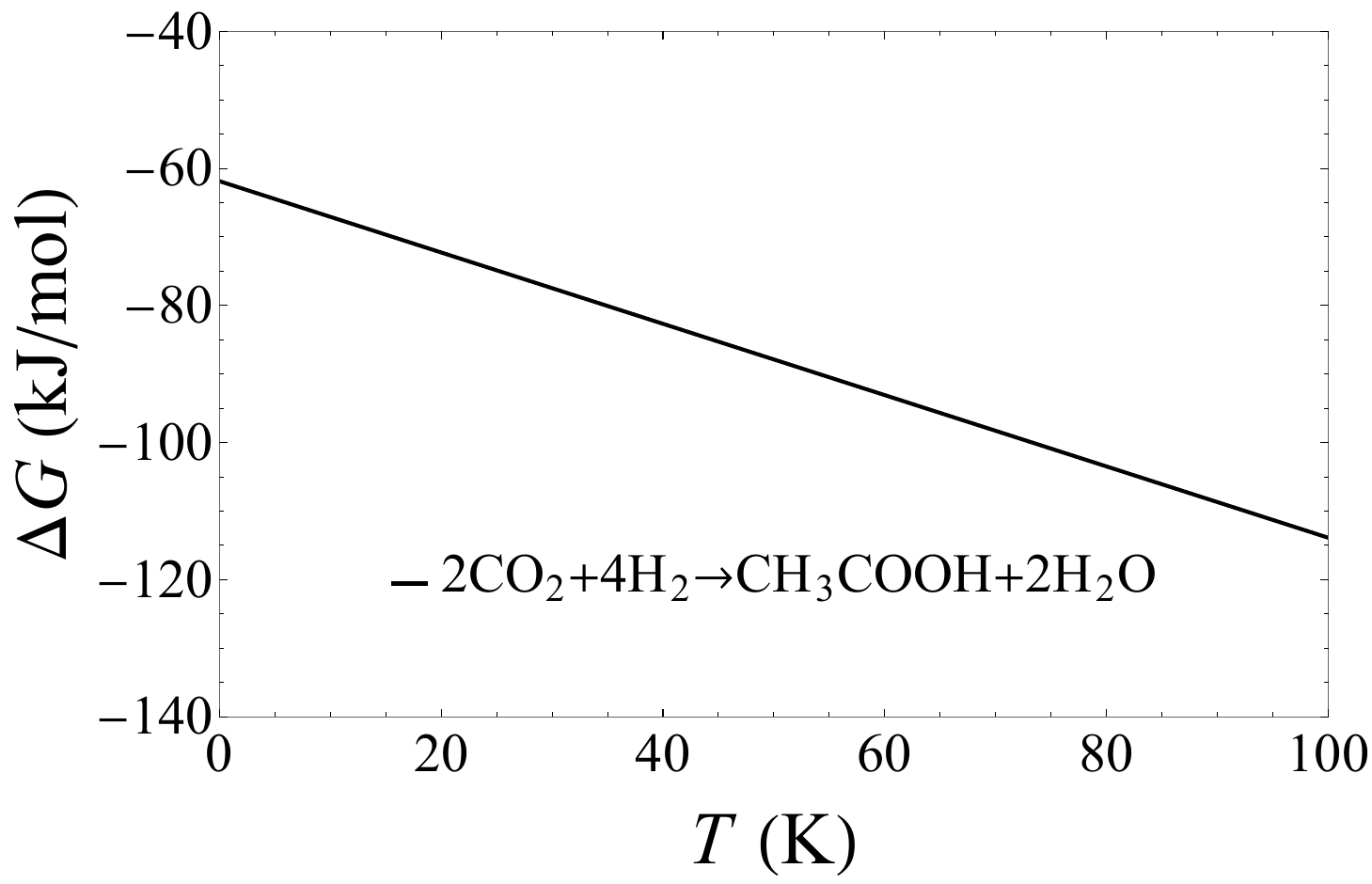}
\caption{The Gibbs free energy released from the synthesis of acetic acid.}
\label{fig:deltaG2}
\end{figure}

In Fig. \ref{fig:deltaG1} and \ref{fig:deltaG2} , we show the Gibbs free energy released from the synthesis of methane (Fig. \ref{fig:deltaG1}) and acetic acid (Fig. \ref{fig:deltaG2}) in the molecular cloud environment. The change of Gibbs free energy is about $\rm -60 \sim -370~kJ/mol$. A negative value of Gibbs free energy is a spontaneous chemical reaction that releases energy. In Ref.  \cite{Kral1998}, the authors measured the free energy values for four methanogens. They found that the minimum energy to sustain the growth of methanogen on Earth is about $\rm 8 - 15~kcar/mol$ (table I in Ref.~\cite{Kral1998}). Following Ref. \cite{McKay2005}, we take an eigenvalue $\rm 10~kcar/mol$ (i.e. $\rm 42~kJ/mol$) as the minimum Gibbs free energy required to maintain the survival of methane bacteria. From Fig. \ref{fig:deltaG1} and \ref{fig:deltaG2}, we can see that the Gibbs free energies released by the reactions are energetically acceptable.

Normally, life forms require less energy at low temperatures and low-pressure environments. However, low pressure represents low number density, which profoundly influences the rate of chemical reactions. The density of acetylene is relatively low, and carbon monoxide is much denser. However, whether the reaction of acetylene and carbon monoxide can provide enough energy depends on the strength of metabolic activity at such low temperatures. Methanogenic and acetogenic life attached to solid carbon dioxide does not have the problem of insufficient raw materials. Solid carbon dioxide can provide sufficient energy and carbon sources for metabolic activities. In addition, there may be other bioenergetics mechanisms as a supplement, such as the mechanism driven by cosmic ray ionization \cite{cosmic-ray-bioenergetics}. Different bioenergetics mechanisms may work together. However, the processes discussed here have unique advantages in fixing carbon by converting carbon elements to organic molecules for life in molecular clouds using methane as an intermediate material.

In this model, the progeny of such methanogenic life would thrive in any solar system locale conducive to its existence, including Europa, Titan, and Mars, which possess dense carbon dioxide atmospheres. There has been an ongoing and intense debate about the detection of methane in Mars' atmosphere. As detailed in Ref. \cite{Formisano}, the authors have reported the discovery of methane in the Martian atmosphere, with a global average mixing ratio of approximately $\rm 10 \pm 5$ parts per billion by volume, as measured by the Planetary Fourier Spectrometer aboard the Mars Express spacecraft. Comparable findings are presented in Refs. \cite{Krasnopolsky,Mumma}. In Ref. \cite{Atreya}, the researchers delve into both abiotic and biotic sources that could supply methane to Mars. However, other orbital observations have failed to detect methane in the Martian atmosphere, as noted in Ref. \cite{Korablev}. This issue still requires further research.

\subsection{A possible distinguishing signal}

The consumption and production of carbon compounds through metabolic activities may significantly affect the distribution of these molecules. It could potentially serve as a trace signal of life within molecular clouds. According to this hypothesis, molecular cloud life uses solid carbon dioxide (carbon monoxide or $\rm C_2H_2$) as a stable carbon source, which causes significant methane or acetic acid production. Consequently, it is reasonable to anticipate that the methane (or acetic acid) distribution within molecular clouds would mirror that of solid carbon dioxide. As the protostars formed in a molecular cloud and the life forms therein potentially concentrate correspondingly, this distribution consistency would be more significant. If such distribution consistency does not exist, it would be a great challenge for our model.

\begin{figure}[htbp]
\centering
\includegraphics[width=0.5\textwidth]{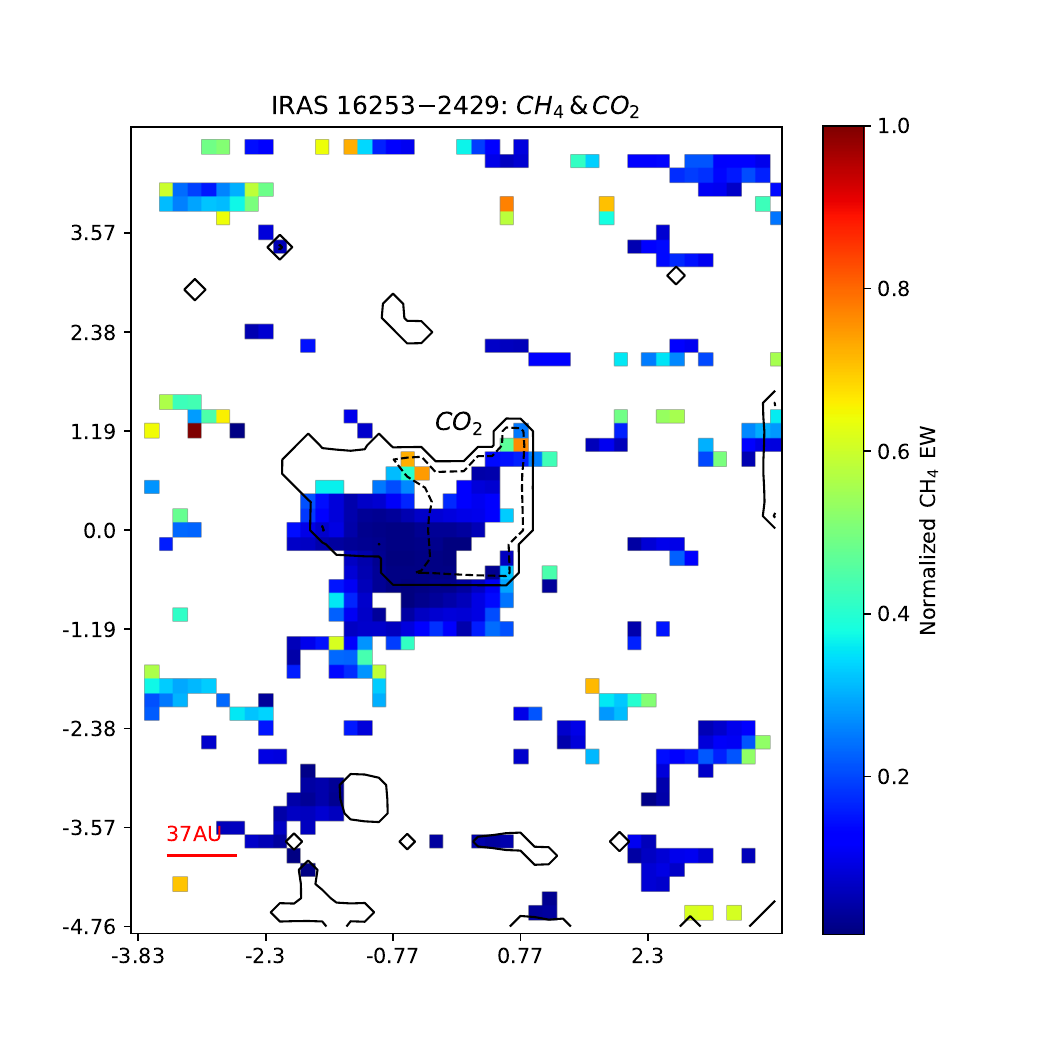}
\caption{The distribution  comparison between carbon dioxide and methane in IRAS16253-2429. The contour denotes the normalized equivalent width of $\rm CO_2$ with 0.1 (line) and 0.7 (dish line). The data comes from Ref. \cite{lei2024}}
\label{fig:IRAS16253}
\end{figure}

Some studies have indicated that carbon dioxide and methane are distributed within the spatial region surrounding two protostars: IRAS 16253-2429 \cite{Narang2023} and IRAS 23385+6053 \cite{Beuther2023}. After the first version of this paper, we further analyzed the data from JWST. Our preliminary analysis indicates that the distribution consistency of carbon dioxide and methane does indeed exist in the region of protostars IRAS16253-2429 \cite{lei2024}. The comparative distribution of carbon dioxide and methane in the protostellar system IRAS16253-2429 is depicted in Fig. \ref{fig:IRAS16253} using the data presented in Ref. \cite{lei2024}. From this figure, it is easy to see that the distribution of carbon dioxide and methane closely aligns with each other.

This distribution consistency has a possible astrochemical origin as it consists of the "Classical" dark-cloud chemistry model \cite{Herbst,Caselli}. The "Classical" dark-cloud chemistry works in an environment of low temperature and gas density,  where ion-molecule reactions dominate the carbon chemistry \cite{Herbst}. As shown in Fig. 6 of Ref. \cite{Caselli}, carbon dioxide and methane are distributed at the outer edge of the pre-stellar core according to the current observations, laboratory work, and modeling.
Whether all or part of methane comes from methanogenic life is a primary research topic for our plans.

\begin{figure}[htbp]
\centering
\includegraphics[width=0.5\textwidth]{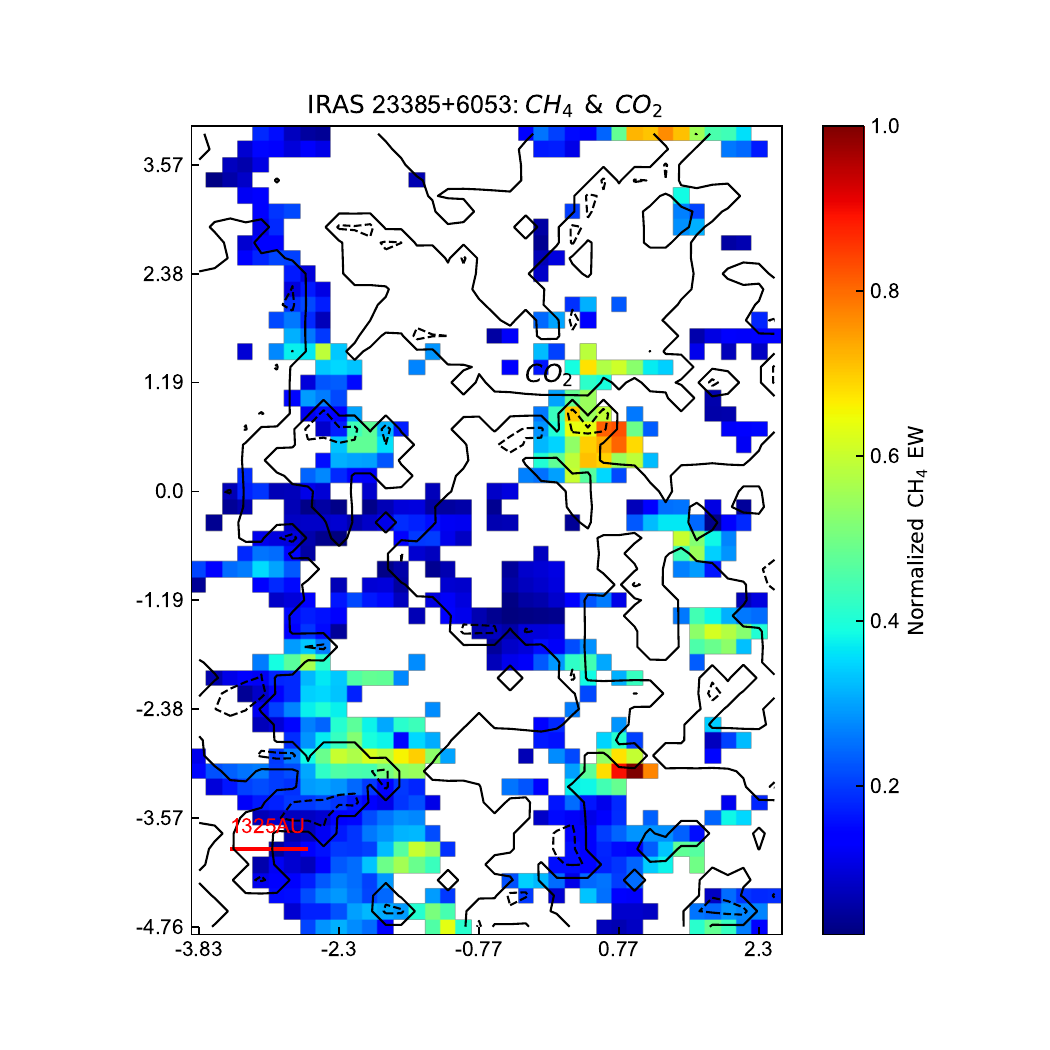}
\caption{The distribution  comparison between carbon dioxide and methane in IRAS23385+6053. The contour denotes the normalized equivalent width of $\rm CO_2$ with 0.1 (line) and 0.7 (dish line). The data comes from Ref. \cite{lei2024}}
\label{fig:IRAS23385}
\end{figure}

For protostars IRAS23385+6053, the distribution consistency is not very good as shown in Fig. \ref{fig:IRAS23385}.  It may be due to the instability induced by the accretion and turbulence processes of IRAS23385+6053 \cite{(Molinari,Fontani,Wolf-Chase,Francis}.

The same conclusion also applies to the distribution of carbon monoxide and $\rm C_2H_2$. We will further investigate whether the distribution of these molecules is consistent with methane in the future.

\section{The relationship with LUCA}

If there were methanogenic or acetogenic life in the pre-solar nebula and then fell on the early Earth, LUCA would be one type of methanogenic or acetogenic life. Since acetogenic bacteria are usually considered a type of bacteria rather than archaea, LUCA may be more inclined towards methanogenic life. The three-domain tree of life presented by ribosomal RNA \cite{ribosomal-RNA} depicted that LUCA is the last common ancestor of archaea, eukaryotes and bacteria. Then acetogenic life could still be a candidate for LUCA.
Coincidentally, the primitive atmosphere of early Earth was rich in carbon monoxide and carbon dioxide\cite{KASTING,KASTING2}. These carbon oxides in the primitive atmosphere may have originated from volcanic eruptions, but depend on the oxidation state of the upper mantle \cite{Holland,Miller1974}. In addition, the impact of comets and asteroids may provide a large amount of $\rm CO$~\cite{DiSanti,KASTING}.
It ensures the continued survival of methanogenic (acetogenic) life on  Earth. Some previous studies \cite{Pierce,Rother} have demonstrated that certain acetogenic and methanogenic Archaea utilized carbon monoxide as a carbon source during the early stages of Earth's development.

The most important model prediction is that these molecular clouds life may be the predecessor of Earth's life in space. More generally, LUCA is one type of life form that uses $\rm H_2$ ($\rm CO$ or $\rm CO_2$) as an electron donor (acceptor) in the chemical reactions of generating energy. In general, this model provides a perfect origination of LUCA. There is already some tentative evidence about the probability that  LUCA is one type of methanogenic or acetogenic life. Some researches indicate that early lifeforms invented methanogenesis in the cooler temperature zones that led to the rise of LUCA after acclimatization to the hydrothermal vent environment by identifying the minimum genome of LUCA \cite{mat}. It has been reported that LUCA's genes point to methanogenic and acetogenic roots by surveying nearly two thousand genomes of modern microbes \cite{Madeline}. Despite a few challenging studies \cite{Berkemer}, research on geological evidence and phylogenomic reconstructions supports this thesis \cite{Preiner,Xavier}.

It should be noted that there is no suggestion here that life originated in molecular clouds. We believe that molecular cloud life may have derived from other planets in the Milky Way or planets in the solar's predecessor star system \cite{nebula-relay}.
\\
\section{Summary}

This paper explores the possibility of methanogenic and acetogenic life in molecular clouds. The calculations demonstrate that the reaction of carbon monoxide, carbon dioxide, or acetylene with hydrogen molecules releases sufficient Gibbs free energy to ensure the survival of molecular cloud life. Solid $\rm CO_2$ may be the primary energy and carbon source because of its high local density. As the second most abundant molecule, the reaction of carbon monoxide is also a good option for providing energy. Additionally, we propose a potential distinguishing signal that could either support or disprove this model. We also consider the possibility that these methanogenic and acetogenic life could have been LUCA's interstellar predecessor. Even now, their descendants and fossils may still widespread in the solar system.

\acknowledgments
We thank Dr. Lei Lei, Yan Sun and Bing-Yu Su for their generous help in perfecting this article.
This work is supported by the National Natural Science Foundation of China (Grants No. 12373002, 11773075) and the Youth Innovation Promotion Association of Chinese
Academy of Sciences (Grant No. 2016288).




\renewcommand{\refname}{References}

\end{document}